\documentclass{elsart3}

\usepackage{times}

\usepackage{graphicx}

\usepackage{amssymb}
\newcommand{\vect}[1]{\mbox{\boldmath $#1$}} 
\begin{document}

\begin{frontmatter}


\title{Phase Diagram for Self-assembly of Amphiphilic Molecule C$_{12}$E$_{6}$ 
by Dissipative Particle Dynamics Simulation}
\author[ia1,ia2]{Hiroaki Nakamura\corauthref{cor1}}
\corauth[cor1]{Corresponding Author:}
\ead{nakamura@tcsc.nifs.ac.jp}
\author[ia1]{Yuichi Tamura}
\address[ia1]{Theory and Computer Simulation Center, National Institute for Fusion Science, 
322-6 Oroshi-cho, Toki, Gifu 509-5292, JAPAN}
\address[ia2]{Department of Fusion Science, School of Physical Sciences, 
The Graduate University for Advanced Studies, 322-6 Oroshi-cho, Toki, Gifu 509-5292, JAPAN}

\begin{abstract}
In a previous study, dissipative particle dynamics simulation was used to qualitatively 
clarify the phase diagram of the amphiphilic molecule 
hexaethylene glycol dodecyl ether (C$_{12}$E$_{6}$).
In the present study, the hydrophilicity dependence of
the phase structure was clarified qualitatively by varying 
the interaction potential between hydrophilic molecules and water molecules
in a dissipative particle dynamics (DPD) simulation using the Jury model.
By varying the coefficient of the interaction potential $x$ between hydrophilic beads and water molecules as 
$x=-20, 0, 10,$ and $20,$ at a dimensionless temperature of $T=0.5$
and a concentration of amphiphilic molecules in water of $\phi=50\%,$ 
the phase structures grew to lamellar ($x=-20$), hexagonal ($x=0$), and micellar 
($x=10$) phases. For $x=20,$ phase separation occurs between
hydrophilic beads and water molecules.
\end{abstract}

\end{frontmatter}

\section{Introduction}\label{sec1} 
 
The phase structure of amphiphilic molecules has been extensively investigated as 
a typical example of soft matter physics.
For the present study, we selected hexaethylene glycol dodecyl ether (C$_{12}$E$_{6}$), 
a popular surfactant in water that has
various self-assembled structures.

\begin{figure}[bp]
\begin{center}
\includegraphics[height=2cm]{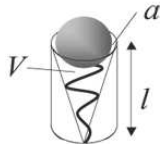}
\end{center}
\caption{Schematic diagram of packing 
parameter\cite{76Israelachvili,92Israelachvili}.
A gray ball and a twisting black line are used to denote the hydrophilic 
and hydrophobic parts, respectively, of an amphiphilic molecule. 
The packing parameter $p \equiv V/Sl$ controls 
the shape of the aggregates. Here, the parameter $V$ is the  
volume occupied by the hydrophobic tail, 
$S$ denotes the sectional area of a hydrophilic group, 
and $l$ is the ``maximum effective length" of the hydrophobic tail. 
\label{packing}
}
\end{figure}

The phase structure of C$_{12}$E$_{6}$ was investigated by Mitchell\cite{83Mitchell} in 1983.
In recent years, phase diagrams at equilibrium, as well as
non-equilibrium and steady-state conditions have been investigated (see Ref. \cite{97Kato}).
Israelachvili proposed the packing parameter as a means of clarifying the relationship 
between macroscopic structure and 
microscopic molecular shape\cite{76Israelachvili,92Israelachvili}.
The packing parameter $p$ is the ratio of the volume $V$ occupied 
by the hydrophobic tail to the product of the sectional 
area of a hydrophilic group $S$ and the ``maximum effective length ($l$)" of the hydrophobic tail  
(see Fig. \ref{packing}).
Spherical micelles are expected when $p \le 1/3$. 
When $1/3 \le p \le 1/2,$ cylinders are expected, and 
for $p\sim 1,$ bilayers should form.

The concept of the packing parameter is intuitive and acceptable.
However, calculating the packing parameter is very difficult, even by 
computer simulation, because it is almost impossible 
to derive macroscopic phase structure at the microscopic level
by simulation, using techniques such as molecular dynamics (MD) simulation, for example.
In order to overcome the gap between macroscopic behavior and microscopic motion, 
dissipative particle dynamics (DPD) simulation has been
proposed as a new mesoscopic motion simulation technique\cite{92Hooger,97Groot,98Groot,01Groot}.
The DPD algorithm might be considered as one of 
the coarse-grained methods of molecular dynamics (MD) simulation.

In 1999, using an empirical method, Jury {\it et al.} succeeded in 
the DPD simulation of the smectic mesophase 
of a simple amphiphilic molecule system with water 
solvent\cite{99Jury}.
Their minimal model (herein referred to  as the Jury model), 
which is composed of rigid AB
dimers in a solvent composed of W monomers,
was shown to be proper for the presentation of the phase diagram of surfactant 
hexaethylene glycol dodecyl ether (${\rm C}_{12}{\rm E}_6$) and water 
(${\rm H}_2 {\rm O}$)\cite{83Mitchell,99Jury}.
In addition, one of the present authors, revealed the dynamical processes of 
the self-organization of one smectic mesophase using the modified Jury model\cite{04Nakamura},
where AB dimer is flexible.

Since some of the information about the interaction potential between particles
is neglected or simplified in DPD simulation, we need to select the dominant
interaction potential for the mesoscopic structure formation.
Since we do not have sufficient experimental data 
for the interaction potentials, defining the interaction parameters 
in DPD simulation becomes difficult. 

The present paper reports an examination of the dependence of
macroscopic phase structure on hydrophilicity by
varying the interaction potential between hydrophilic molecules and water molecules
in DPD simulation as a first step  toward clarifying 
the relationship between interaction potentials and the macroscopic structure
(Section 3).
By strengthening hydrophilicity, water-particles penetrate  closer to the hydrophilic 
heads (A), and therefore the heads go apart from each other. Moreover, the length of 
AB dimer becomes larger, because a repulsive force between water (W) and hydrophobic tail (B) 
becomes stronger. 
In this way, it is expected that $p$ can be varied and that macroscopic structure 
deforms.

In Section 3, we compare the simulation results and 
the experiments for   C$_{12}$E$_{6}$ and C$_{12}$E$_{8}$.
We also discuss about  another interaction potential, that is, the head-head (A-A) interaction.

\section{Simulation Method}\label{secSim}

\subsection*{DPD Algorithm}\label{DPD}
In the present study, we used the DPD model and 
algorithm\cite{97Groot,99Jury,04Nakamura}.
According to the ordinary DPD model, 
all atoms are coarse-grained to particles 
of the same mass.
The total number of particles is defined as $N.$
The position and velocity vectors of particle $i, 
(i=1,\cdots, N),$ are indicated by $\vect{r}_i$ 
and $\vect{v}_i$, respectively. 
Particle $i$ moves according to the following 
equation of motion, where all physical quantities 
are made dimensionless in order to 
facilitate handling in actual simulation.
\begin{eqnarray}
\frac{d \vect{r}_{i} }{d t} &=& \vect{v}_i ,      
\label{eq1} \\
\frac{d \vect{v}_{i} }{d t} &=& \sum_{j (\ne i)}^{N} 
\vect{F}_{ij},   
\label{eq2}
\end{eqnarray}
where particle $i$ interacts with another  
particle, $j$, according to the total force, $\vect{F}_{ij}$,
which is comprised of four forces as follows: 
\begin{equation}
\vect{F}_{ij} =
\vect{F}_{i j}^{\rm C} + \vect{F}_{i j}^{\rm R} 
+\vect{F}_{i j}^{\rm D}  
+\vect{F}_{i j}^{\rm B}. \label{eq3}
\end{equation}
In Eq.~\ref{eq3},
$\vect{F}^{\rm C}_{ij}$ is a conservative force 
derived from a potential 
exerted on particle $i$ by particle $j$, 
$\vect{F}^{\rm D}_{ij}$ and 
$\vect{F}^{\rm R}_{ij}$ are the dissipative and 
random forces between particles $i$ and $j$, respectively. 
Furthermore, neighboring particles on the same amphiphilic 
molecule are bound by the bond-stretching force 
$\vect{F}^{\rm B}_{ij}$. 
The conservative force $\vect{F}^{\rm C}_{ij}$ 
has the following form:
\begin{equation}
\vect{F}^{\rm C}_{ij} \equiv   
  \left\{
    \begin{array}{ll}
      a_{i j} (1-r_{i j}) \vect{n}_{ij} 
         & \mbox{if  }  r_{i j}<1,\\
    0 & \mbox{if  } r_{i j} \ge 1,     
    \end{array}
  \right.
\label{eq.fc}
\end{equation}
where $ \vect{r}_{ij} \equiv \vect{r}_{i}-\vect{r}_{j}, 
        r_{ij} = |\vect{r}_{ij}|,$ and 
$\vect{n}_{ij} \equiv \frac{\vect{r}_{ij} }{|{r}_{ij}|}.$
For computational convenience, we adopted 
a cut-off distance of unit length.
The conservative force 
$\vect{F}^{\rm C}_{ij}$ is assumed to be truncated 
beyond this cutoff. Coefficients $a_{ij}$ 
denote the coupling 
constants between particles $i$ and $j$.

Espa\~nol and Warren proposed 
the following simple form of the random 
and dissipative forces \cite{95Espanol}:
\begin{eqnarray}
{\vect{F}}_{i j}^{\rm R} &=& \sigma  \omega (r_{i j})  
\vect{n}_{ij} 
\frac{ \zeta_{ij }}{\sqrt{\Delta t}},
\label{eq.fr} \\
{\vect{F}}_{i j}^{\rm D} &=& -\frac{\sigma^2}{2T}
\omega (r_{i j})^2
 \left(  \vect{v}_{i j}\cdot \vect{n}_{ij} 
 \right) \vect{n}_{i j} , \label{eq.fd}
\end{eqnarray}
where $\vect{v}_{i j}=\vect{v}_{i}-\vect{v}_{j}$ and $\zeta_{ij}$ is a Gaussian random valuable 
with zero mean and unit variance that is chosen independently 
for each pair $(i, j)$ of interacting 
particles at each time-step and $\zeta_{ij} = \zeta_{ji}.$ 
The strength of the dissipative forces is 
determined by the dimensionless parameter $\sigma$. 
The parameter $\Delta t$ 
is the dimensionless time-interval used to integrate
the equation of motion.
Here, the function $\omega$ is defined 
by\cite{97Groot,95Espanol}:
\begin{equation}
\omega(r) = \left\{ 
                \begin{array}{@{\,}ll}
               1-r & \mbox{if  }  r< 1,\\
               0 & \mbox{if  } r \ge 1. 
                 \end{array}
              \right.
\end{equation}

Finally, we use the following form as  
the bond-stretching force:
\begin{equation}
\vect{F}_{i j}^{\rm B} = - a_{\rm B} \omega(r_{ij})
  \vect{n}_{ij},
\label{eq.phi}
\end{equation}
where $a_{\rm B} $ is the potential energy coefficient.

\begin{table}
\begin{center}
\large
\begin{tabular}{|c|ccc|}   \hline
$a_{ij}$  & ~~W~~ & ~~A~~ & ~~B~~ \\ \hline
\ \ \ W \ \ \ & 25 &  $x$    & 50 \\
\ \ \ A \ \ \ &  $x$  & 25 & 30 \\
\ \ \ B \ \ \ & 50 & 30 & 25  \\ \hline
\end{tabular} 
\end{center}
\caption{\footnotesize \label{aij}Table of coefficients $a_{ij}$ depending 
on particle type for particles $i$ and $j$, where W is a water particle, 
A is a hydrophilic particle, and B is a hydrophobic particle. 
By varying the coefficient $x$ between A and W particles as
$x = -20, 0, 10$ and $20,$ the dependence of the phase structure on
the hydrophilicity is clarified.  }
\end{table}

\subsection*{Simulation Model and Parameters}\label{model}
We used the modified Jury model molecule for a dimer composed of
a hydrophilic particle (A) and a hydrophobic particle (B)\cite{99Jury,04Nakamura}.
In addition, water molecules were modeled as coarse-grained particles (W).
The masses of all particles were assumed to be unity.
The number density of particles $\rho$ was set to $\rho=6$.
The number of modeled amphiphilic molecules AB was $N_{\rm AB}$, where
the number of water molecules was $N_{\rm W}.$
The total number of particles $N\equiv 2N_{\rm AB} + N_{\rm W}$ 
was fixed to $N=10000.$
The simulation box was set to cubic.
The dimensionless length of the box $L$ was 
\begin{equation}
L=\left(\frac{N}{ \rho } \right)^{ \frac{1}{3}} 
\sim 11.85631.   \label{eq.L}
\end{equation}
In simulation, we used a periodic boundary condition.
The interaction coefficient $a_{ij} $ in Eq.~\ref{eq.fc} is 
given in Table~\ref{aij}. 
In order to clarify the dependence of the phase structure on
molecular shape, we varied the coefficient $x$ between A and W particles as
$x=-20, 0, 10,$ and $20.$ 
When the coefficient $x$ is positive, the conservative force between A and W
becomes repulsive. On the other hand, negative $x$ gives the attractive force between A and W.

The coefficient of the bond-stretching interaction potential $a_{\rm B}$  
is adopted as $a_{\rm B} = 100.$
We set the dimensionless time-interval of one step to $\Delta t = 0.05.$

As the initial configuration, 
all of the particles were located randomly.
The velocity of each particle was distributed
so as to satisfy a Maxwell distribution with dimensionless
temperature $T.$ 
The dimensionless strength of dissipative forces was 
$\sigma = 3.3541 \sqrt{T}.$
During the simulation, we set $T=0.5$ and 
$\phi = 50\%.$

\section{Simulation Results and Discussions}\label{results}
\begin{figure}[htbp]
\begin{center}
\includegraphics[width=6cm]{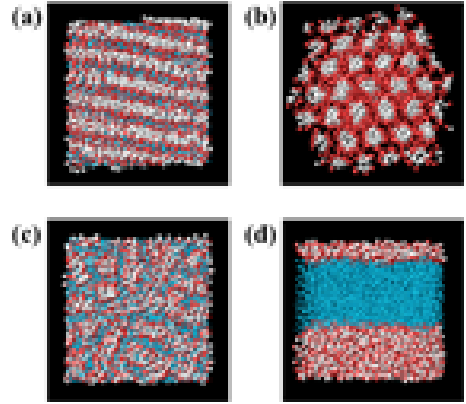}
\caption{\label{st1}Formed structures for each potential coefficient,
$x=-20 (a), 0 (b), 10 (c),$ and $20 (d).$ Each structure is shown
in Table \ref{phase}. Red and white beads denote 
hydrophilic (A) and hydrophobic
molecules (B), respectively. 
Blue beads represent groups of water molecules (W).
We set $T=0.5$ and $\phi=50\%$ during simulation.}
\end{center}
\end{figure}

We demonstrated the dependence of macroscopic phase structure on hydrophilicity by
varying the A-W interaction potential coefficient $x$.
By varying the coefficient of the interaction potential $x$ as 
$x=-20, 0, 10,$ and $20,$
the phase structures became lamellar ($x=-20$), hexagonal ($x=0$), and micellar 
($x=10$) phases. For $x=20,$ phase separation occurs between
hydrophilic beads and water molecules.
The structure for each $x$ is 
shown in Fig. \ref{st1} and summarized in Table \ref{phase}.
This figure shows that the  packing parameter $p$ becomes smaller from $p\sim 1$ (lamellar phase) 
to $p \sim 1/3$ (micellar phase),
when the interaction coefficient $x$ becomes larger (i.e. less hydrophilic).
 (When $x=20$,   phase separation appears. 
  In this case,  the packing parameter $p$ cannot be used to clarify the phase structure.)
Thus, we could demonstrate that 
the packing parameter can be varied indirectly by changing the hydrophilicity.

\begin{figure}[htbp]
\begin{center}
\includegraphics[width=7cm]{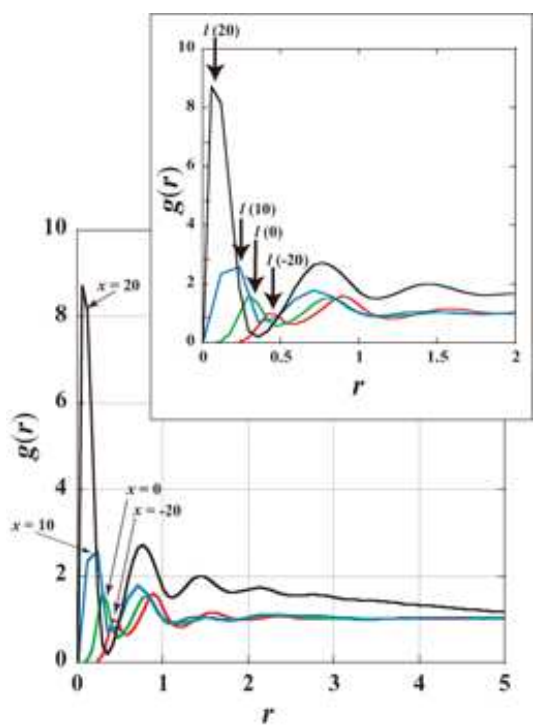}
\end{center}
\caption{\label{gr}Solute particle radial distribution function $g(r)$ 
vs. distance between two particles $r$ for $x=-20, 0, 10 $ and $20$.
The function $g(r)$ is the sum of the A-A radial distribution function,
the B-B radial distribution function, and the A-B radial distribution function.
The first peak $l(x)$ of each curve corresponds to the length of AB dimer. 
}
\end{figure}

Next, we discuss the dependence of the shape of AB dimer on varying the A-W interaction.
In order to obtain the information on the molecular shape, 
we plot the radial distribution function of the solute particles $g(r)$ for each $x$ in Fig. \ref{gr}.
To be exact, we comment the definition of the $g(r)$; 
$g(r)$ is the sum of  A-A,  B-B, and A-B radial distribution functions.
We marked the first  peaks for each $x$ in the upper-right frame in Fig. \ref{gr}. 
The bond-stretching interaction in AB dimer (Eq. \ref{eq.phi}) is the most attractive force
among all interaction forces in the present model (Table \ref{aij}).
Therefore, the distance  between A and B in an intra-molecule corresponds 
to the first peak $l(x)$ of $g(r)$.
From Fig. \ref{gr}, it is found that $l$ becomes  larger, 
as  the parameter  $x$ becomes smaller 
(i.e. the A-W interaction becomes more hydrophilic).
On the other hand, Fig. \ref{st1} showed 
that  $p$ becomes larger, when  $x$ becomes smaller.
Therefore, it is found that the a conical AB dimer with  the head particle (A) attached to a short tail (B) 
forms spherical micelles for large $x$ and that AB dimer varies its shape from cone to
cylinder by increasing  tail's length $l$, as $x$ becomes smaller.

Last, we comment on amphiphilic molecule experiments. 
The phase diagram of C$_{12}$E$_{6}$ is different from that of  C$_{12}$E$_{8},$ because 
the hydrophilic head of C$_{12}$E$_{6}$ is shorter than that of C$_{12}$E$_{8}$.
It is known\cite{83Mitchell} that the lamellar phase region  in the  phase diagram of  C$_{12}$E$_{8}$ 
is narrower than that of  C$_{12}$E$_{6}$. 
Moreover, the hexagonal phase region in the phase diagram of C$_{12}$E$_{8}$  
is larger than that of  C$_{12}$E$_{6}$.
In the present model, C$_{12}$E$_{8}$  corresponds to   smaller $x$ (i.e. more hydrophilic) 
than  C$_{12}$E$_{6}$.
From our simulation, it is found that
the hexagonal phase trends to  the lamellar phase,
as the $x$ becomes smaller.
Therefore, it is expected that the hexagonal phase region of the diagram for smaller $x$
becomes smaller than that for  larger $x$.
This prediction by simulation contradicts the experimental fact.
This contradiction derives its origin from the fact that we adopted only
the head-water interaction parameter $x$ as a variable  
to clarify the macroscopic phase.
It might be seen intuitively reasonable to adopt only the head-water interaction  
as  a descriptor for distinguishing  C$_{12}$E$_{6}$ vs. C$_{12}$E$_{8}$.
However, the difference between C$_{12}$E$_{6}$ and C$_{12}$E$_{8}$ is not only
the strength of the head-water interaction but also that of the head-head interaction,
by which  the packing parameter can be  controlled  directly. 
As the result, 
we found that the head-head interaction  dominates the structure formation process of
C$_{12}$E$_{n}$ series more than the head-water interaction.



\begin{table}[htbp]
\begin{center}
\begin{tabular}{|c||c|}   \hline
$x$     & ${\rm Formed Structures}$  \\ \hline \hline
~~-20~~ & ~~L$_{\rm \alpha}$~~       \\ \hline
 ~~0~~  & ~~H$_{1}$~~                \\ \hline
~~10~~  & ~~L$_{1}$~~                \\ \hline
~~20~~  & Phase separation           \\  \hline
\end{tabular} 
\end{center}
\caption{\label{phase}Table of formed structures for each $x$.
Lamellar, hexagonal, and micelles phases are indicated 
as L$_{\rm \alpha},$  H$_{1},$
and L$_{1},$ respectively.
For $x=20,$ AB molecules and W molecules are separated, as shown in
Fig. \ref{st1}. }
\end{table}


\section*{Acknowledgments}  
This research was partially supported by the Ministry of Education, Culture, Sports, 
Science and Technology of Japan, Grant-in-Aid for Scientific Research (C), 2003, No.15607019.

\bibliography{reference2003}

\begin{thebibliography}{10}
\expandafter\ifx\csname url\endcsname\relax
  \def\url#1{\texttt{#1}}\fi
\expandafter\ifx\csname urlprefix\endcsname\relax\def\urlprefix{URL }\fi

\bibitem{83Mitchell}
D.~J. Mitchell, G.~J.~T. Tiddy, L.~Waring, {P}hase behaviour of polyoxyethylene
  surfactants with water, J. Chem. Soc., Faraday Trans. 1 79 (1983) 975.

\bibitem{97Kato}
T.~Kato, Microstructure of nonionic surfactants, Structure-Performance
  Relationships in Surfactants 72 (1997) 325--357.

\bibitem{76Israelachvili}
D.~J.~M. J.~Israelachvili, B.~W. Ninham, J. Chem. Soc. Faraday Trans. I 72
  (1976) 1525.

\bibitem{92Israelachvili}
J.~Israelachvili, {I}ntermolecular and {S}urface {F}orces, 2nd Edition, Academi
  Press, London, 1992.

\bibitem{92Hooger}
P.~J. Hoogerbrugge, J.~M. V.~A. Koelman, {S}imulating microscopic hydrocynamic
  phenomena with dissipative particle dynamics, Europhys. Lett. 19 (1992) 155.

\bibitem{97Groot}
R.~D. Groot, P.~B. Warren, {D}issipative {P}article {D}ynamics: {B}ridging the
  gap between atomistic and mesoscopic simulation, J. Chem. Phys. 107 (1997)
  4423.

\bibitem{98Groot}
R.~D. Groot, T.~J. Madden, {D}ynamic simulation of diblock copolymer microphase
  separation, J. Chem. Phys. 108 (1998) 8713.

\bibitem{01Groot}
R.~D. Groot, K.~L. Rabone, {M}esoscopic simulation of cell membrane damage,
  morphology change and rupture by nonionics surfactants, Biophys. J. 81 (2001)
  725.

\bibitem{99Jury}
S.~Jury, P.~Bladon, M.~Cates, S.~Krishna, M.~Hagen, N.~Ruddock, P.~Warren,
  {S}imulation of amphiphilic mesophases using dissipative particle dynamics,
  Phys. Chem. Chem. Phys. 1 (1999) 2051.

\bibitem{04Nakamura}
H.~Nakamura, Mol. Simu. (2004) in press.

\bibitem{95Espanol}
P.~Espa{\~n}ol, P.~Warren, {S}tatistical mechanics of dissipative particle
  dynamics, Europhys. Lett. 30 (1995) 191.

\end{thebibliography}

\end{document}